# Towards Sub-micrometer High Aspect Ratio X-ray Gratings by Atomic Layer Deposition of Iridium

Joan Vila-Comamala[a,b,*], Lucia Romano[a,b,c], Vitaliy Guzenko[a], Matias Kagias[a,b], Marco Stampanoni[a,b], Konstantins Jefimovs[a,b]

[a]*Institute for Biomedical Engineering, University and ETH Zürich, 8092 Zürich, Switzerland*
[b]*Paul Scherrer Institut, 5232 Villigen PSI, Switzerland*
[c]*Department of Physics and CNR-IMM, University of Catania, 95023 Catania, Italy*

## Abstract

X-ray grating interferometry is an excellent technique for X-ray phase contrast imaging and X-ray wavefront sensing with applications in materials science, biology and medical diagnosis. Among other requirements, the method depends on the availability of highly X-ray absorbing metallic gratings. Here, we report on the fabrication and characterization of high aspect ratio iridium gratings with a period of one micrometer and a depth of 30 µm combining deep reactive ion etching of silicon and atomic layer deposition of iridium. The implementation of such structures can greatly enhance the sensitivity of grating-based X-ray phase contrast imaging and thus, expand further its broad range of applications.

[*]Corresponding author
  *Email address:* joan.vila-comamala@psi.ch (Joan Vila-Comamala)

## 1. Introduction

X-ray grating interferometry[1, 2, 3] is a prominent method for X-ray wavefront sensing[4, 5] and X-ray phase contrast imaging with applications in materials science, biology and medical diagnosis[6, 7, 8]. The technique crucially depends on the availability of highly X-ray absorbing metallic gratings. For absorbing typical X-ray photon energies of 20 keV and above, thicknesses of at least 25–30 µm of a high atomic number element, such as gold, are required. On the other hand, typical X-ray grating periods range from 15 down to a few micrometers, and thus, high aspect ratio structures are necessary for most grating-based X-ray phase contrast imaging setups. Recently, X-ray phase constrat imaging applications using even smaller grating periods have been proposed[7, 9] but currently used fabrication methods can not routinely produce gratings with periods of one micrometer and below. To date, X-ray absorption gratings are commonly fabricated combining optical or X-ray lithography with gold electroplating in silicon templates[10, 11] or polymer resist molds[12]. However, such fabrication techniques are not easily transferred for the production of sub-micrometer high aspect ratio structures. Here, we propose and demonstrate the production of high aspect ratio ($> 50$) iridium gratings with a period of one micrometer and a depth of 30 µm combining deep reactive ion etching (RIE) of silicon and atomic layer deposition (ALD) of iridium[13]. Until now, ALD processes have been successfully applied to conformally deposit oxide layers on high aspect ($> 180$) nanostructures[14, 15] and ALD metal coatings have also been proven very effective to fabricate high resolution diffractive X-ray optics in the sub-100 nm structure size range[16, 17]. In this work, we performed a thorough optimization of the ALD recipe to extend the conformal metal coating to trench depths of tens of micrometers while keeping the one micrometer grating periodicity. After scanning electron microscopy inspection, the fabricated gratings were successfully implemented and characterized in a laboratory X-ray phase contrast imaging setup.

## 2. Materials and Experimental Methods

### 2.1. Grating-based Phase Contrast X-ray Imaging

The interaction of X-rays with an object can be described by introducing a complex refractive index, $n = 1 - \delta + i\beta$, in which $\beta$ and $\delta$ respectively account for the absorption and phase shift experimented by X-ray wavefield due to the specimen[18]. For biological soft tissues with small density differences, the phase shift refractive index variation $\delta$ can be up to three orders of magnitude larger than its absorption counterpart $\beta$, thus allowing X-ray phase contrast imaging to deliver images with much higher contrast in comparison to those obtained by conventional X-ray absorption imaging. Among different existing X-ray phase contrast imaging techniques, X-ray grating interferometry obtains excellent contrast and quantitative information of the sample under investigation[19, 20]. Grating-based phase contrast imaging can be both performed in synchrotron radiation facilities[2] and using incoherent laboratory X-ray sources[3]. In the latter case, the setup typically requires the use of two highly absorbing X-ray gratings, $G_0$ and $G_2$, and phase-shifting grating, $G_1$, as schematically represented in Fig. 1. The combination of $G_0$ and $G_1$ gratings creates an X-ray intensity modulation at a given downstream distance (Talbot distance) from the $G_1$ phase-shifting grating. The $G_2$ grating is necessary to detect the intensity modulation when the pixel size of the X-ray detector is larger than the period of this X-ray intensity modulation. After inserting the sample near the $G_1$ phase-shifting grating, the X-ray intensity modulation is modified. This change allows the detection of the sample differential phase contrast image by step-scanning one of the three gratings in small steps covering a range of one or several grating periods while recording consecutively the transmission X-ray intensity images[2]. The exact geometry and grating periods requirements for an X-ray grating interferometer setup can be calculated using the formulas reported elsewhere[21].

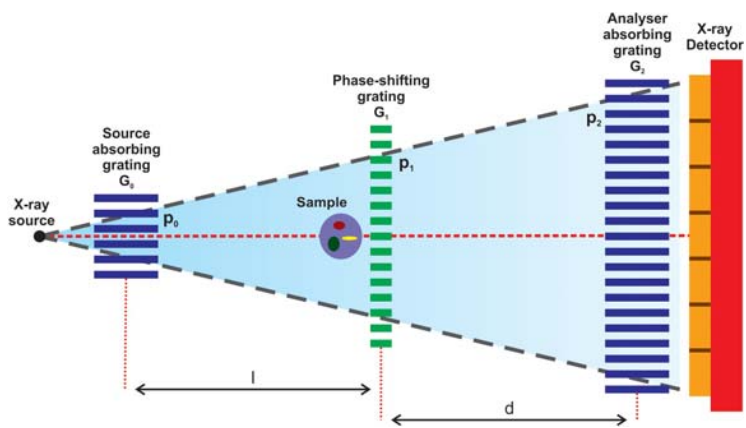

Figure 1: Schematics of laboratory grating-based phase contrast X-ray imaging setup composed of two highly absorbing X-ray gratings, $G_0$ and $G_2$, and a phase-shifting X-ray grating, $G_1$ (not to scale).

The fabricated X-ray gratings were tested in a laboratory setup using a Hamamatsu L10101 X-ray microsource with a tungsten target and an expected focused source size of 5–10 μm. The X-ray microsource high voltage was set to 40 kV and an electron current of 0.200 mA was employed. The X-ray images were acquired using a scintillator-based sCMOS camera with pixel size of 22 μm (1024x1024 pixels) from Photonic Science Ltd. (UK). The gratings were mounted on motorized stages with nanometer positioning accuracy for alignment and scanning simplicity. The gratings periods $p_0$, $p_1$ and $p_2$ were chosen to be one micrometer. Thus, two absorbing gratings made of iridium were mounted in our laboratory X-ray phase contrast imaging setup as $G_0$ and $G_2$. The phase-shifting $G_1$ grating was made of silicon and it had a thickness of 25 μm, thus producing the required phase shift of π rad for an X-ray energy of approximately 20 keV.

*2.2. Fabrication of the X-ray Gratings*

Phase contrast imaging X-ray gratings require thin silicon supporting substrates to minimize the X-ray absorption when typical photon energies below 30 keV are employed. Nevertheless, the handling of very thin substrates can be challenging because of their fragility and finally, 4-inch silicon double side polished wafers with a thickness of 250 μm were used to fabricate the X-ray gratings reported here. The schematics of the fabrication process is depicted in Fig. 2. Before the lithography step, the silicon wafers were coated with a 100 nm thick layer of chromium by electron beam evaporation. After that, the grating pattern was produced using conventional UV photolithography on positive photoresist layer (MicroChem Corp. S1805) using a Karl Suss Mask Aligner MA6 in vacuum contact mode. After exposure and development of the photoresist, the pattern was further transferred into the chromium hard mask by a RIE $Cl_2$/$CO_2$ based process. Then, the photoresist residuals were removed by immersion in acetone and isopropyl alcohol. Using the patterned chromium layer as hard mask, the high aspect ratio silicon structures were produced by a $SF_6$ / $C_4F_8$ based deep RIE, also commonly know as Bosch process, using an inductively coupled plasma (ICP) Plasmalab 100 system from Oxford Instruments Plasma Technology (UK). The high aspect ratio and high resolution structures were achieved after a careful fine tuning of all process parameters such as the gas flows, chamber pressure and radiofrequency powers. As schematically suggested in Fig. 2(c), the photolithography and deep silicon RIE processes were adjusted to produce a 2 µm period silicon grating with a duty cycle of 0.25. In this way, the one micrometer iridium grating was only produced after a conformal metal coating at both sides of the silicon trenches. Since silicon is a very low absorbing material in comparison to iridium, the gaps and the silicon lines of the grating can be regarded as equivalent for the X-ray wavefield that will only be sensitive to effective iridium grating period. In the past, such an approach has been successfully demonstrated for the production of high resolution X-ray diffractive optics[16, 17]. Finally, the last step in grating fabrication was the conformal deposition of iridium by ALD using iridium acetylacetonate, $Ir(acac)_3$, and oxygen gas, $O_2$, as precursors[13], as depicted in Fig. 2(d). In principle, this deposition method can be used to coat conformally any surface structure by repeatedly supplying two complementary reactant vapors in alternating pulses. Because the chemical reactions are forced to happen entirely on the surface and they are self-limited by the amount of precursor reactant that can be adsorb by the surface, the technique is well-suited for a film growth with almost atomic monolayer accuracy[22, 23]. During the investigations presented here, a Picosun™ R-200 Advanced ALD tool with the capability of delivering a plasma-enhanced oxygen precursor was used. This ALD system has a large reaction chamber suitable for up to 8-inch wafers and it has 6 separate inlets allowing for several precursors to be supplied. The inlets are constantly flushed with a flow of nitrogen, even during the ALD cycle steps, to prevent the growth of materials and the likely clogging of the precursor inlet orifices. As a result, the reactor chamber also requires constant vacuum pumping to prevent an excessive pressure increase, which is kept at around 10 mTorr. Since the precursor species are supplied by mixing the precursor vapor with inlet flow of nitrogen, the constant vacuum pumping of the reactor chamber is artificially limiting the precursor exposure of the surface being coated. However, the system can be operated using a Picoflow™ diffusion enhancer operation mode which combines the temporary closure of the vaccuum pumping valve during the precursor delivery pulse of every ALD cycle with the addition of a stainless steel lid to reduce the reactor chamber size. Using this operation mode, the effective dosing time is increased and the diffusion of the precursor species is promoted. Concerning the precursor species, the Ir(acac)$_3$, that is a powder at room temperature, is kept at high temperature (195 deg C) and low pressure, so that it sublimates inside its reservoir. The precursor is delivered by shortly opening the container valves and mixing the Ir(acac)$_3$ vapor with nitrogen flow through the line. Therefore, the Ir(acac)$_3$ vapor pressure limits the total amount of precursor that can be delivered for a single ALD pulse to the reactor chamber. The O$_2$ precursor can be both delivered as gas for an exclusively thermal reaction or through a plasma generator for a plasma-enhanced ALD process with a typical radiofrequency power of 2000 W.

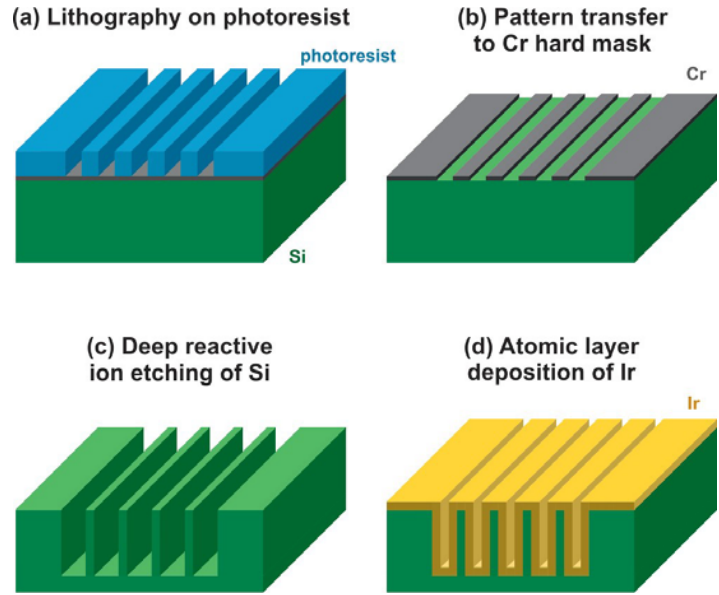

Figure 2: Fabrication steps of the X-ray gratings by combining deep reactive ion etching of silicon and atomic layer deposition of iridium. After the photolithography step (a), the grating pattern is transferred into the chromium hard mask, (b), used during the silicon deep reactive ion etching, (c). The final atomic layer deposition step, (d), coats the silicon trenches with a conformal layer of iridium.

## 3. Results and Discussion

### 3.1. Optimization of X-ray Grating Fabrication

The fabrication procedure of the X-ray gratings reported here relies on combining two well-established techniques – silicon deep RIE and ALD of iridium – that required an exhaustive tailoring for our particular purpose. On the one hand, the optimization of the X-ray grating fabrication started with the fine-tuning of the deep reactive ion etching by balancing the etching and passivation times to achieve the required depth of 30 µm, vertical sidewalls with a 2 µm periodicity and the duty cycle of 0.25. The details of the deep RIE recipe are summarized in Tab. 1 and its results are shown in the scanning electron micrograph of Fig. 3(a). It can be seen that all the requirements are met and that the targeted 500 nm silicon lines with a periodicity of 2 µm and depth of 30 µm have been achieved.

Table 1: Detailed parameters of the silicon deep reactive ion etching recipe. An ICP Plasmalab 100 system from Oxford Instruments Plasma Technology (UK) was used.

|  | Etching Step | Passivation Step |
| --- | --- | --- |
| Temperature [deg C] | 0 | |
| Number of Cycles | 350 | |
| Presurre [mtorr] | 20 | |
| $SF_6$ Flow [sccm] | 100 | 5 |
| $C_4F_8$ Flow [sccm] | 5 | 100 |
| ICP Power | 600 | 600 |
| RF Power | 30 | 20 |
| Step Time [s] | 3.0 | 3.0 |

On the other hand, the optimization of the ALD recipe was both approached from a theoretical point of view and from the experimental considerations directly derived from the operational aspects of the Picosun™ R-200 Advanced ALD tool reported in Sec. 2.2. From a theoretical standpoint, ALD is expected to coat conformally any surface if sufficient precursor exposure and dosing times are supplied. Nevertheless, a complete step coverage will strongly depend on the particular aspect ratio of the structures under consideration[22, 24]. For example, and from a kinetic theoretical model, it has been shown that a hole with aspect ratio 43 is expected to require a pre-cursor exposure of about 3000 times larger than a flat surface[22]. Therefore, high aspect ratio structures will require maximizing the precursor exposure of the surface by increasing the precursor partial pressure (i.e. concentration) in the reactor chamber or at least, by increasing the dosing time to promote the precursor diffusion and to ensure the precursor saturation of the surface being coated. Taking this into account, the use of silicon gratings with a periodicity of 2 µm and a duty cycle 0.25 is clearly more advantageous than trying to completely fill smaller trenches of 500 nm.

After a few experimental trials, it was confirmed that the essential key for a successful ALD conformal coating was to deliver enough Ir(acac)$_3$ precursor and to promote its diffusion into the silicon trenches by increasing the ALD cycle step times and by employing the Picoflow™ operation mode of the system. The complete parameter details of the ALD recipe that delivered the highest conformal coating to the 30 µm deep silicon trenches are contained in Tab. 2. This optimized iridium ALD recipe started with a deposition of a thin (∼ 10 nm) layer of $Al_2O_3$ to promote the nucleation of the iridium layer on the silicon surface. After that, the actual iridium deposition was divided into two separate processes: (1) a short deposition using oxygen plasma-enhanced recipe to obtain a uniform nucleation; and (2) a very long deposition using a thermal oxygen process to prioritize the conformality of the iridium coating. It is reported that the use of a plasma precursor is beneficial to obatin a nucleation of smaller grain which will result in a much smoother layer as the deposition progresses. However, it has also been reported that for long depositions the plasma can lead to unconformal deposition as recombination problems can occurs for high aspect ration structures, that is, the precursor ion species may react before reaching the bottom region of the trench[23]. The results of the ALD recipe can be observed in the scanning electron micrograph of Fig. 3(b) and they demonstrate the conformal coating of the 30 µm deep silicon trenches. Figures 3(c) and (d) compare the iridium thickness obtain at the top and the bottom of the silicon template and Fig. 3(d) shows a larger portion of the fabricated grating.

Table 2: Detailed parameters of the iridium atomic layer deposition recipe for conformal coating of high aspect ratio silicon trenches. The recipe is divided in three separate processes: a deposition of a thin layer of Al$_2$O$_3$ to ensure the iridium growth; the nucleation of iridium using O$_2$ plasma-based recipe to obtain smoother iridium layer; the deposition of the required iridium thickness using thermal O$_2$-based recipe to prioritize a conformal coating of the silicon.

|  | Al$_2$O$_3$ Deposition | | Ir Nucleation | | Ir Deposition | |
| --- | --- | --- | --- | --- | --- | --- |
|  | 1st ALD Pulse | 2nd ALD Pulse | 1st ALD Pulse | 2nd ALD Pulse | 1st ALD Pulse | 2nd ALD Pulse |
| Temperature [deg C] | 370 | | 370 | | 370 | |
| Number of Cycles | 300 | | 800 | | 11000 | |
| Thickness [nm] | 10 | | 30 | | 480 | |
| Presurre [mTorr] | 12 | 12 | 12 | 12 | 20 | 12 |
| Precursor | TMA | H$_2$O | Ir(acac)$_3$ | O$_2$ plasma | Ir(acac)$_3$ | O$_2$ gas |
| Line Gas Flow [sccm] | 150.0 | 150.0 | 150.0 | 60 | 150.0 | 100.0 |
| Pulse Time [s] | 0.5 | 0.5 | 1.0 | 6.0 | 4.0 | 10.0 |
| Purge Time [s] | 6.0 | 6.0 | 6.0 | 6.0 | 35.0 | 35.0 |
| Picoflow™ Operation Mode | No | No | No | No | Yes, 25.0 s | Yes, 25.0 s |
| Total Cycle Time [s] | 13 | | 19 | | 120 | |
| Total Deposition Time | < 1 hour | | ~ 4 hours | | ~ 15 days | |

The current ALD recipe using the Picosun™ R-200 Advanced ALD tool was able to deliver a conformal coating of iridium for 3 grating substrate pieces of 2 × 2 cm$^2$ with a single ALD run. However during the experimental trials, it was found that the iridium deposition was not completely conformal at the central regions of a full 4-inch silicon wafer with a ~ 7 × 7 cm$^2$ grating patterned area. In this latter case, it is suspected that the total amount of Ir(acac)$_3$ precursor supplied to the chamber was not sufficient to saturate the larger area of the fully patterned 4-inch silicon wafer. This result clearly points the importance of operational aspects of the particular ALD system in use. It is expected that with a dedicated ALD system that would be able to deliver the Ir(acac)$_3$ precursor from more than one reservoir at the same time would allow the coating of larger grating areas and it would probably decrease substantially the very long ALD pulse times required by the equipment used for investigations reported here.

*3.2. Performance of the X-ray Gratings*

The X-ray absorbing iridium gratings were characterized at our laboratory X-ray phase imaging setup. A picture of the actual system is shown in Fig. 4(a). The setup was arranged to employ the 9th and 13th Talbot order configurations with a grating separation distances of 3.7 and 5.2 cm, respectively. The fabricated gratings were mounted and aligned using the motorized positioning systems. During the tests, a sample consisting of polystyrene microspheres with a diameter of 700 µm was used. The differential phase contrast images of the spheres at the 9th and 13th Talbot order configurations are respectively shown in Fig. 4(b) and (c). Comparing the two images, the highest sensitivity is achieved by the 13th Talbot order as it is theoretically expected[21]. The main advantage of having smaller periods is the reduction of the grating separation distance to obtain a similar sensitivity. For example, when using 3 µm grating periods a separation distance between gratings of 16.3 cm would be required for a similar sensitivity to the one achieved by setup reported here. In addition, using shorter grating separation distances is very beneficial in terms of photon flux when using laboratory X-ray mi-crosources. The measured fringe grating visibility, that is an evaluation of the performance of the X-ray phase imaging setup, was found to be 11.5% and 12.1% for the 9th and 13th Talbot orders. The visibility histograms are shown in the inset plots of Fig. 4(b) and (c). Finally, it can be observed that the phase contrast images are clean of defects, which in turn is an indication of the excellent quality of the fabricated one micrometer period gratings.

## 4. Conclusions

In summary, we have demonstrated the fabrication and characterization of high aspect ratio iridium gratings with a period of one micron and a depth of 30 µm for X-ray phase contrast imaging. The combination of deep RIE of silicon and atomic layer deposition of iridium has been proven as an effective fabrication method to produce high quality X-ray absorbing gratings. The technique has the potential of producing even smaller period gratings over larger areas ($> 5 \times 5$ cm$^2$) if used together with new emerging lithography technologies such as Displacement Talbot Lithography[25] that have the capability of producing sub-micrometer grating patterns over large areas (4 and 8-inch wafers). We believe that the proposed method can be used to provide high quality gratings that will greatly enhance the sensitivity of grating-based X-ray phase contrast imaging and broaden its range of applications. In addition, the technique could be easily transferred for the production of other microcomponents and optical devices.

**Acknowledgements**

The authors would like to thank Dr. Z. Wang and Dr. C. Arboleda from the Paul Scherrer Institut (Switzerland) for assistance during the phase contrast X-ray imaging experiments. This work has been partially funded by the ERC-2012-SRG 310005-PhaseX grant and by the Swiss National Science Foundation, SNSF Grant Number 159263.

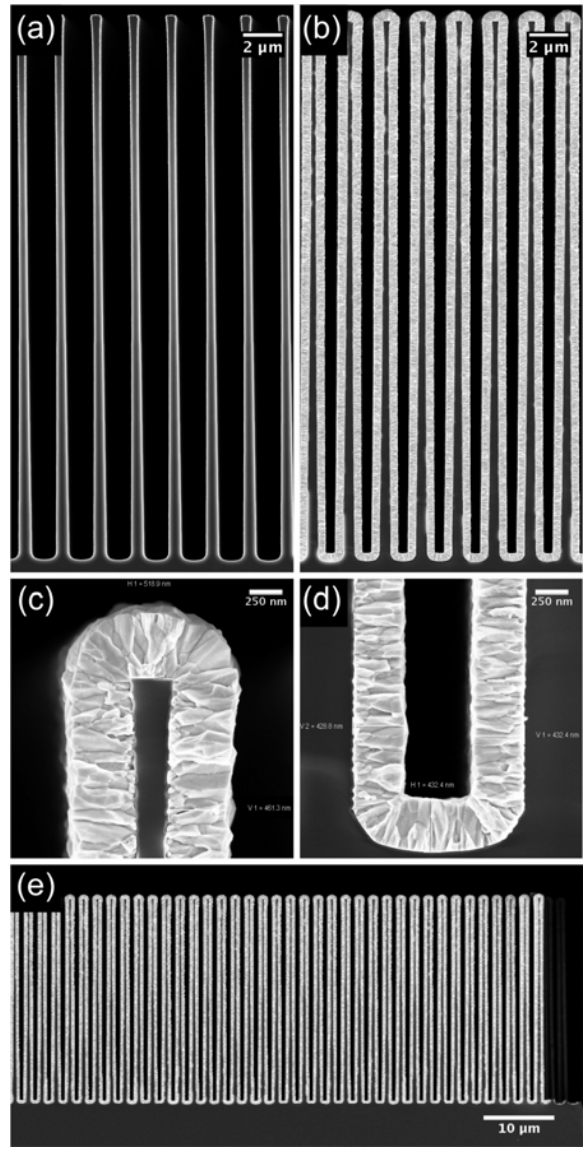

Figure 3: (a) Scanning electron micrograph of the silicon grating fabricated by deep reactive ion etching with a periodicity of 2 µm, a depth of 30 µm and a duty cycle of 0.25. (b) One micrometer iridium grating after the atomic layer deposition process. (c) and (d) demonstrate the obtained conformality by comparing the thickness of iridium at the top and bottom of the silicon trenches. (e) (a) Low magnification scanning electron micrograph of the one micrometer iridium X-ray grating.

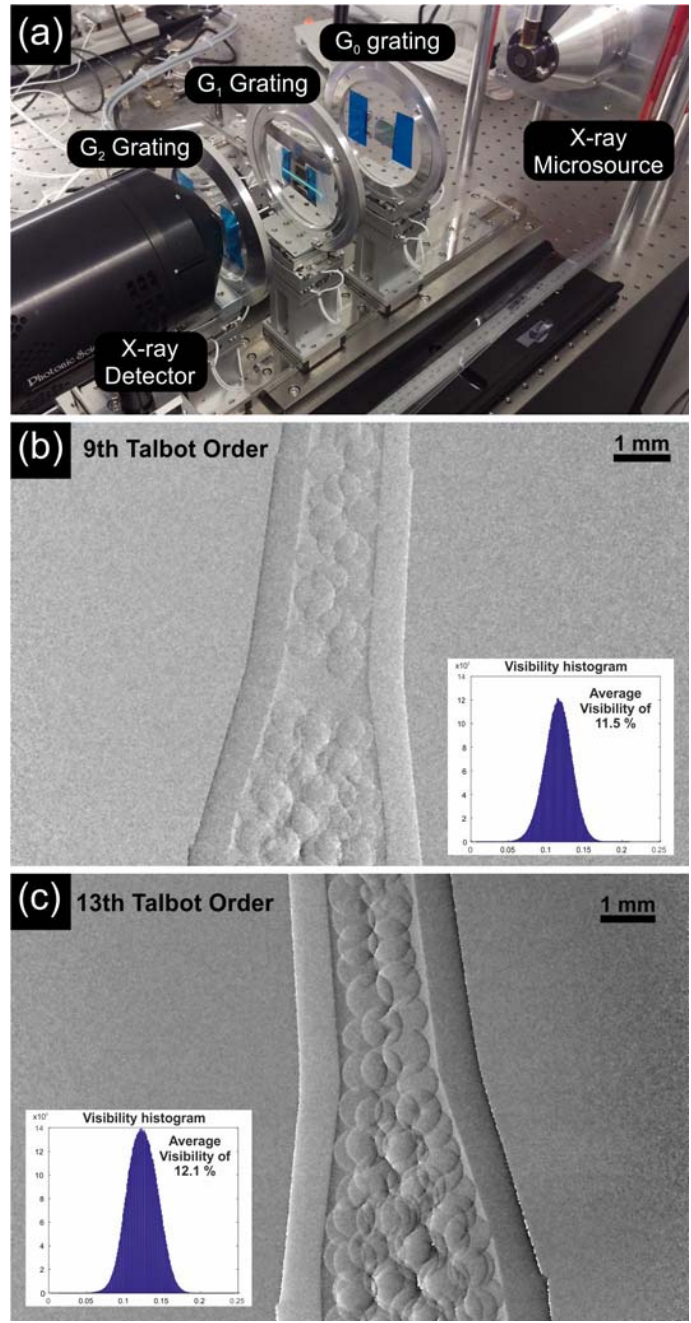

Figure 4: (a) Picture of the laboratory X-ray phase imaging setup with the fabricated one micrometer period iridium gratings mounted as $G_0$ and $G_2$. (b) and (c) display the differential phase contrast images of a sample made of 700 μm silicon oxide spheres obtained at the 9th and 13rd Talbot order $2r_0$ configurations. The inset plots shows the histogram of the visibility obtained, that is a direct evaluation parameter of the grating-based X-ray phase contrast imaging setup.